\documentclass[
twocolumn,
aps,
prd,
longbibliography,
showkeys,
amsmath,
amssymb,
floatfix,
superscriptaddress,
]{revtex4-2}

\usepackage{import}

\usepackage{color,graphics,epsfig,rotating}
\usepackage{graphicx}
\usepackage{dcolumn}
\usepackage{bm}
\usepackage[dvipsnames]{xcolor}

\usepackage{placeins}
\usepackage{mathrsfs}
\usepackage{lmodern}
\usepackage{notoccite}

\usepackage[breaklinks=true]{hyperref}
\usepackage{breakcites}
\usepackage[capitalise]{cleveref}
\usepackage{amsmath}
\usepackage{grffile}
\usepackage{xspace}
\usepackage{xifthen}
\usepackage{soul}
\usepackage{placeins} 
\usepackage{orcidlink}
\usepackage{natbib}
\usepackage{xfrac}
\usepackage{graphicx}
\usepackage{listings}
\usepackage{xcolor}

\usepackage{gensymb}

\usepackage[normalem]{ulem}

\newcommand{\pd}{\partial}

\newcommand*{\Eint}{\Delta E_\mathrm{int}}

\begin{document}

{
\onecolumngrid
Notice: This manuscript has been coauthored by UT-Battelle, LLC, under Contract No. DE-AC0500OR22725 with
the U.S. Department of Energy. The United States Government retains and the publisher, by accepting the article for
publication, acknowledges that the United States Government retains a non-exclusive, paid-up, irrevocable, world-wide
license to publish or reproduce the published form of this manuscript, or allow others to do so, for the United States
Government purposes. The Department of Energy will provide public access to these results of federally sponsored
research in accordance with the DOE Public Access Plan (\href{http://energy.gov/downloads/doe-public-access-plan}{http://energy.gov/downloads/doe-public-access-plan}).
}

\title{Inversion of Dislocation-Impurity Interactions in \texorpdfstring{$\alpha$}{α}-Fe under Magnetic State Changes}

\author{Franco Moitzi\,\orcidlink{0000-0003-0558-7966}}
\email{moitzi.francopeter@gmail.com}
\affiliation{Christian Doppler Laboratory for Digital material design guidelines for mitigation of alloy embrittlement, Materials Center Leoben Forschung GmbH, Vordernberger Stra{\ss}e 12, A-8700 Leoben, Austria}
\author{Lorenz Romaner\,\orcidlink{0000-0003-4764-130X}}
\affiliation{Chair of Physical Metallurgy and Metallic Materials, Department of Materials Science, 
             Montanuniversit{\"a}t Leoben, Roseggerstra{\ss}e 12, A-8700 Leoben, Austria}
\author{Andrei V. Ruban\,\orcidlink{0000-0002-3880-0965}}
\affiliation{Materials Center Leoben Forschung GmbH, Vordernberger Stra{\ss}e 12, A-8700 Leoben, Austria}
\affiliation{Department of Materials Science and Engineering, Royal
         Institute of Technology, 10044 Stockholm, Sweden}
\author{Swarnava Ghosh\,\orcidlink{0000-0003-3800-5264}}
\affiliation{National Center for Computational Science, Oak Ridge National Laboratory, Oak Ridge, Tennessee 37831, USA}
\author{Markus Eisenbach\,\orcidlink{0000-0001-8805-8327}}
\affiliation{National Center for Computational Science, Oak Ridge National Laboratory, Oak Ridge, Tennessee 37831, USA}
\author{Oleg E. Peil\,\orcidlink{0000-0001-9828-4483}}
\affiliation{Materials Center Leoben Forschung GmbH, Vordernberger Stra{\ss}e 12, A-8700 Leoben, Austria}

\date{\today}

\begin{abstract}

In this work, we 
investigate the dislocation-impurity interaction 
energies and their profiles for 
various \textit{3d} elements \textemdash V, Cr, Mn, Cu, Ni, and Co \textemdash in and 
around $1/2\langle111\rangle$ screw dislocations in $\alpha$-Fe using \textit{ab initio} methods.
We consider the ferromagnetic and paramagnetic states, with the latter being modeled through 
both the disordered local moment model and a spin-wave approach.
Our findings reveal that (1) magnetic effects are large compared to 
size misfit effects of substitutional impurities, 
and (2) dislocation-impurity interactions are
dependent on the magnetic state of the
matrix and thermal lattice expansion. In particular, Cu 
changes from core-attractive in the 
ferromagnetic state to repulsive in the paramagnetic state.

\end{abstract}


\maketitle

Dislocation-impurity interactions are crucial in the plastic deformation of metals, as dislocations 
often dictate mechanical behavior. Alloying enables tailoring of these interactions by controlling impurity type and concentration, allowing for targeted tuning of material properties.

In body-centered cubic (bcc) metals at moderate temperatures, deformation is dominated 
by screw dislocation glide via thermally activated kink-pair mechanisms~\cite{Weinberger2013}. 
Yet, trends in yield strength can be predicted based on 
isolated $1/2\langle111\rangle$ screw dislocations in pure bcc metals \cite{Vitek1970} or 
bcc alloys \cite{Romaner2010,Romaner2014,Romaner2017,Odbadrakh2016,Casillas2023}. 
To offer a more tractable approach to strength in alloys, 
models have been introduced that describe the 
solute-induced strengthening or softening through 
dislocation–impurity interaction energies~\cite{Rao2021,Rao2023,Maresca2020}.

However, the \textit{ab initio} description of the technologically important $\alpha$-Fe
is challenging due to the complex magnetic states, which include 
ferromagnetism (FM) at low temperatures and paramagnetism (PM) above the Curie-Temperature $T_C$ in the
form of disordered moments. At the dislocation core, different local atomic coordination, volumetric changes, and 
magnetic coupling exert a significant influence on interatomic interactions, which can be seen from 
the different magnetic moments at the core~\cite{Dezerald2014, Clouet2008, Ventelon2015, Casillas2020}. 
Alloying $\alpha$-Fe with other \textit{3d} magnetic elements introduces 
additional complex interactions that cannot be fully explained by elastic size mismatches 
alone~\cite{Varvenne2017}.

Only a few \textit{ab initio} studies have 
investigated the energy profiles of magnetic impurities around dislocation cores \cite{Odbadrakh2016,Casillas2023}, with the latter also examining the influence of the PM states on 
energy profiles. These studies only focused on Ni and Cr, and as a result, 
they did not discover impurities displaying altered interaction behavior with increasing temperature.

In the present study, we investigate from first principles the impact of a broader range of impurities (V, Cr, Mn, Cu, Ni, and Co) and magnetic states on $1/2\langle111\rangle$ screw dislocations in $\alpha$-Fe alloys.
We find that for almost all elements in question, the interaction energy changes sign when going from the
low-temperature FM to high-temperature PM state, with the strongest effect being observed for Cu.
To check the robustness of our results, we employ several different density-functional theory (DFT) methodologies
and two methods for treating the PM state. 
Furthermore, we also examine how the choice of exchange-correlation (XC) functional 
and the inclusion of lattice relaxations affect the energy profiles.

Calculations are performed using the Vienna Ab-initio Simulation Package (VASP) 
\cite{Kresse1993,Kresse1996a,Kresse1996b} with semicore \textit{p}-states included as valence states, the Exact Muffin-Tin Orbital (EMTO) 
method with full-charge density formalism \cite{Vitos2001a, Vitos1997} (Lyngby version \cite{Ruban2016}) 
\textemdash
which combines Green's function-based 
DFT with the coherent potential approximation (CPA) for total energy computations \textemdash and the 
EMTO-based Locally Self-Consistent Green's Function (ELSGF) 
technique~\cite{Abrikosov1997,Peil2012}. CPA in ELSGF and EMTO 
enable disordered local moment (DLM) approximation
for describing the PM state \cite{Staunton1984,Gyorffy1985}.

As an alternative to DLM, we also use the 
Locally-Self Consistent Multiple Scattering (LSMS) code~\cite{Wang1995,Eisenbach2017,rogers2016overcoming}
for carrying out spin-wave method (SWM)~\cite{Ruban2012} calculations, which requires large supercells and averaging over special planar spin-wave configurations. 

Effects of thermal expansion are taken into account by considering two 
lattice constants: 
$a_{\mathrm{LT}} = 2.85$ \r{A} for low T (LT) and $a_{\mathrm{HT}} = 2.90$ \r{A} for high T (HT),
approximately corresponding to experimental values for 0 K and around the Curie temperature
~\cite{seki2005lattice,Basinski1955}.

Unless specified otherwise, the calculations are done with the local density approximation
(LDA) exchange-correlation (XC) functional of Perdew-Wang~\cite{pw91}.
This choice is motivated by later discussions on differences between LDA and generalized gradient approximation (GGA) results.

A dislocation model with 135 atoms per 1 Burgers ($1\Vec{b}$) vector was used~\cite{Cai2003,Ventelon2015}. 
We compare two setups: 
``Str'', a chain of impurity atoms along $\vec{b}$, and 
``Imp'', a single impurity in Fe along $\vec{b}$. In the CPA-EMTO method, ``Imp'' 
represents the dilute Fe-X alloy, while other ``Imp'' cases use a $4\Vec{b}$-height cell. 
Details are available in the Supplemental Materials (SM).

\begin{figure}[!ht]
   \centering
   \includegraphics{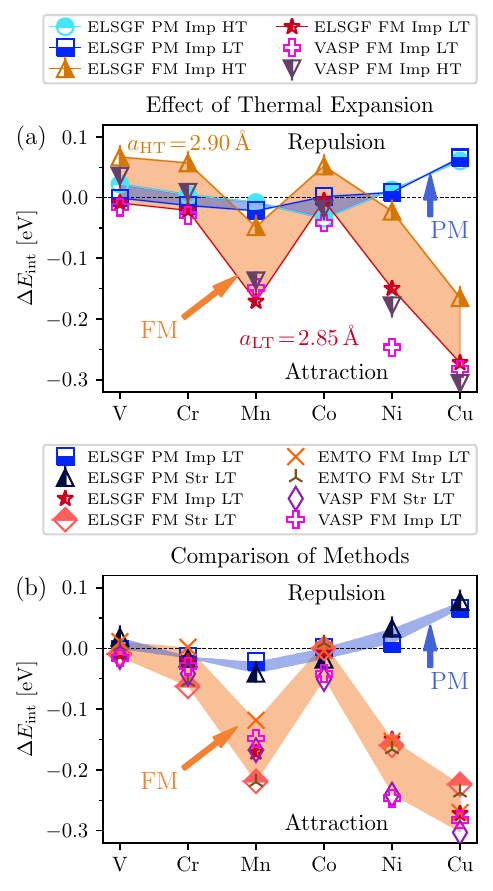}
   \caption{Dislocation-impurity interaction energies to the dislocation core site for \textit{3d}-elements (V, Cr, Mn, Co, Ni, Cu) in bcc Fe. The energies are calculated for both paramagnetic (PM) and ferromagnetic (FM) states using ELSGF, EMTO, and VASP methods. (Str) refers to a string of impurities along the $\Vec{b}$-direction, while (Imp) represents a single impurity atom (or the dilute limit in CPA) surrounded by Fe atoms along the $\Vec{b}$-direction. 
   (a) Effect of thermal expansion through variation of the lattice constant from LT to HT;
   (b) effect of a specific method and setup.}
   \label{fig:fig_01}
\end{figure}

We characterize the dislocation-impurity interaction strength by 
evaluating the relative dislocation-impurity interaction energy,
taken as the difference between the energy at a core site and a reference site far away from 
the core, 
    $\Eint
    = 
    E_{\mathrm{X}\rightarrow\mathrm{Fe}}^{\mathrm{core}}
  - E_{\mathrm{X}\rightarrow\mathrm{Fe}}^{\mathrm{ref}}$,
where $\mathrm{X}\rightarrow\mathrm{Fe}$ denotes an 
impurity $\mathrm{X}$ substituted at an $\mathrm{Fe}$ site.

This interaction energy, also termed solute-dislocation interaction energy, is an 
important quantity in models describing the dislocation 
mobility~\cite{Rao2023} and solute-induced 
effects on strength~\cite{hu2017solute}. It can also be interpreted as a \emph{relative} segregation 
energy, where negative/positive values indicate a preference/aversion for the core site over the 
reference site.

For core sites, we select one of the three equivalent positions surrounding the 
screw dislocation core (first shell in the differential displacement map marked by 
a triangle \cref{fig:fig_04}), while the reference site, positioned as far as 
possible from the core, is located six atomic shells away (rightmost site, within the purple box in \cref{fig:fig_04}).

The relative dislocation-impurity interaction energies, $\Eint$, calculated with 
ELSGF in both FM and PM states and with VASP in the FM state for six different impurities
(V, Cr, Mn, Co, Ni, Cu) and for the two lattice constants are presented in \cref{fig:fig_01}a. 
We consider \( a_\mathrm{LT} \) and \( a_\mathrm{HT} \) for both magnetic states, because,
on the one hand, this allows us to describe the intermediate regime at elevated temperatures
from $\sim 800$ K to the Curie temperature ($T_C = 1043$ K), where the system is still
in the FM state, but significant disorder is already present.
On the other hand, calculations at \( a_\mathrm{LT} \) in the PM state are performed to
check the general lattice-constant dependence of this magnetic state and to get data
for fitting a simple interpolative model described further in the text.

In the FM state at \( a_\mathrm{LT} \),
V, Cr, and Co exhibit positive $\Eint{}$ close to zero, implying weak attraction to the dislocation core. 
In contrast, Mn, Ni, and especially Cu, have a strong tendency to segregate to the core. 
Expanding the lattice from \( a_\mathrm{LT} \) to  \( a_\mathrm{HT} \) shifts energies 
upwards (towards more repulsive values) without altering trends, with a smaller 
shift in VASP ($\sim0.05$ eV) than ELSGF ($\sim0.08$ eV).

As to the PM state, it is characterized by only a marginal volume effect
on $\Eint$. V, Cr, and Co having energies similar to their counterparts
at $a_\mathrm{LT}$ in the FM state. 
Mn, Ni, and Cu, on the other hand, demonstrate a behavior very
different from that in the FM state. Specifically, Mn has nearly zero negative
dislocation-impurity interaction energy, Ni exhibits weak attraction, while Cu becomes a significantly repulsive impurity.
Juxtaposing the most physically relevant cases of the FM state at $a_\mathrm{LT}$
and the PM state at $a_\mathrm{HT}$, we can see that Ni and Cu clearly invert 
the sign of their interaction. For Mn, $\Eint$ does not change sign, but its
absolute value drops dramatically, almost to zero.  V, Cr, and Co are largely unaffected 
by magnetism, responding mainly to lattice constant changes.

Focusing on the FM-PM transition at $a_\mathrm{HT}$, 
the two element groups, (Mn, Ni, Cu) and (V, Cr, Co), show 
opposite trends. For the first group, $\Eint{}$ increases making interactions 
more repulsive. For the second, they become more attractive 
(or, at least, less repulsive).

The trends in $\Eint{}$ described above remain consistent across various impurity configurations and 
computational methods, as shown in \cref{fig:fig_01}b. However, 
important insights can still be gained through detailed comparisons
(1) between the ``Str'' and ``Imp'' setups to 
assess impurity-impurity interactions along the $\Vec{b}$-direction, and (2) within the 
``Imp'' setup, between the ELSGF and EMTO-CPA methods, to evaluate the impact of local 
environmental effects.

Regarding the first aspect, when Mn and Cr are arranged 
in an impurity chain, as opposed to being surrounded by Fe in the $\Vec{b}$-direction, 
they become more attracted to the core, while Cu behaves in the opposite 
way in the FM state. In contrast, no such effect is observed in the PM state.

As to the second aspect, the results from ELSGF and EMTO-CPA coincide for all 
elements for the ``Str'' setup, but not for the ``Imp'' setup, where V, Cr, and
Mn exhibit small discrepancies between the two methods.
The strongest effect is observed for Mn in the FM state at $a_{\mathrm{LT}}$,
where $\Eint{} = -0.11$ eV for EMTO-CPA and $\Eint{} = -0.16$ eV for ELSGF with
the ``Imp'' setup.
This difference can be traced back to the influence of the local environment on
the magnetic moments of Mn, resulting in slightly lower values of the latter in ELSGF
compared to EMTO-CPA.

Notably, elements experiencing the strongest local environment effects, namely, V, Cr, and Mn,
have a tendency to align their magnetic moments antiparallel with respect to the Fe matrix.
For the ``Str'' setup, this can lead to magnetic frustration, if exchange interactions between the
impurities along the dislocation line are also of the antiferromagnetic type.
Such an effect seems to be significant only in case of Mn, for which EMTO and ESLGF yield
a quenched magnetic moment, while VASP and LSMS stabilize magnetic moments of Mn,
but the convergence is difficult, suggesting that there might be multiple metastable states.
In such a situation, non-collinear configurations might become favorable, and we
leave this question to further studies, as these subtle differences in magnetic arrangements
do not alter the overall qualitative picture presented in \cref{fig:fig_01}a.

\begin{figure}
   \centering
   \includegraphics{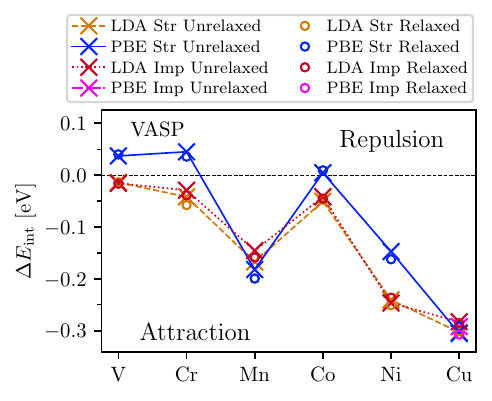}
   \caption{Comparison of dislocation-impurity interaction energies 
   ($\Eint{}$) for \textit{3d}-elements (V, Cr, Mn, Co, Ni, Cu) in bcc Fe evaluated using both PBE and LDA. $\Eint{}$ is obtained with (Relaxed) and without (Unrelaxed) atomic relaxation following impurity introduction.
   (Str) refers to a string of impurities along the $\Vec{b}$-direction ($1b$-cell),
   while (Imp) represents a single impurity atom surrounded by Fe atoms along the $\Vec{b}$-direction 
   in a $4b$-cell
   }
   \label{fig:fig_02}
\end{figure}

Other sources of uncertainty include the choice of XC functional and atomic relaxations. 
To account for this, \cref{fig:fig_02} shows $\Eint{}$ for relaxed and 
unrelaxed impurity configurations, calculated with VASP in the FM state 
at $a_\mathrm{LT}$ using LDA and PBE. Our PBE results for Ni and Cr agree with Refs.~\onlinecite{Casillas2023, Odbadrakh2016}.

PBE just shifts energies of some elements upward compared to LDA, leaving the trends 
qualitatively consistent. While 
gradient corrections might improve accuracy in non-centrosymmetric core region, PBE may overestimate 
magnetic exchange contributions \cite{Wu2006,Park2015,Moitzi2022}. 
LDA's well-known error in determining the equilibrium volume is irrelevant
in our calculations, as the lattice constant is fixed to experimental values.
Hence, we choose LDA to ensure correct description of magnetism.

The effect of relaxations is estimated using VASP by allowing an impurity atomic configuration
to relax and comparing the obtained energy to the unrelaxed one. Here we only mean relaxations after the
impurity substitution, the initial core configuration of pure Fe being fully relaxed
in the FM state. 

As seen in \cref{fig:fig_02}, the relaxation effects are practically negligible
for both LDA and PBE, which has already been reported in
Ref.~\onlinecite{Odbadrakh2016} for Ni and Cr. The outcome is expected because
the size mismatch between Fe and elements under consideration is marginal, and
the impurities do not alter the core configuration.

The equilibrium dislocation structure in the PM state is 
very similar to that in the FM state, with both 
exhibiting the easy-core configuration, which was found by via non-collinear DLM \cite{Casillas2020} 
and SWM \cite{RazumovskiyRomaner2024} calculations. Also for the PM state, we expect, therefore, the relaxation effects 
of impurities to be marginal.

\begin{figure}
   \centering
   \includegraphics{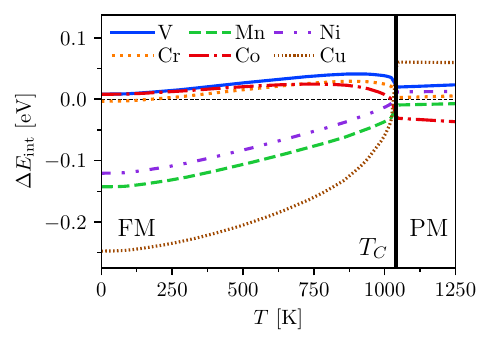}
   \caption{Dislocation-impurity interaction 
   energies as a function of temperature from 0 K to Curie temperature at 1043 K obtained by
   fitting $\Eint(m, a)$ to experimental magnetization $m(T)$ and lattice constant $a(T)$.
   }
   \label{fig:fig_03}
\end{figure}

To illustrate the impact of temperature on the dislocation-impurity interaction
during heat treatment, let us define the interaction energy as a function,
$\Eint(m, a)$, of the reduced magnetization, $m = M / M_{s}$ \textemdash with $M$ being the magnetization
at a given temperature and volume, $M_{s}$ the
saturation magnetization at 0 K \textemdash and of the lattice parameter, 
$a = \left[ a_\mathrm{LT}, a_\mathrm{HT} \right]$.
Considering the results in \cref{fig:fig_01}a as four boundary cases with
$m = \{0, 1\}$ (corresponding to the PM and FM states, respectively) and of the lattice constant, $a = \{a_\mathrm{LT}, a_\mathrm{HT}\}$, for each impurity, we can define $\Eint(m, a)$ as an interpolation
within these bounds. The temperature evolution described in that way is qualitative, as it relies 
on a simplified model for $\Eint(m, a)$ and neglects effects like 
phonon entropy.

We choose a bilinear interpolation:
\begin{equation} 
\Eint(m, a) = c_1 + c_2 m + c_3 a + c_4 a \cdot m, 
\end{equation}
as the simplest model that accounts for the coupling between interatomic interactions and magnetization 
\cite{Gorbatov2013a} (see SM for details). 
Using the experimental values for the lattice constant,
$a(T)$ \cite{seki2005lattice}, and the reduced magnetization, $m(T)$ \cite{Crangle1971}, 
we show the temperature-dependent interaction energies 
$\Eint(T) \equiv \Eint\big( m(T), a(T) \big)$ in \cref{fig:fig_03}, confirming 
earlier conclusions for Mn, Ni, and Cu.

However, the $T$-dependence of $\Eint$ for V, Cr, and Co reveals a non-trivial
behavior: Weak 
attraction/repulsion at LT, followed by an increasing repulsion due to lattice expansion at HT, 
and finally, interaction energies falling back to small values in the PM state

Moreover, this group of elements exhibits a maximum in $\Eint$ at some temperature
below the Curie temperature. This non-monotonic behavior can be traced back to the positive
value of $\Eint(1, a_\mathrm{HT}) - \Eint(0, a_\mathrm{HT})$ (see SM).

\begin{figure}
   \centering
   \includegraphics{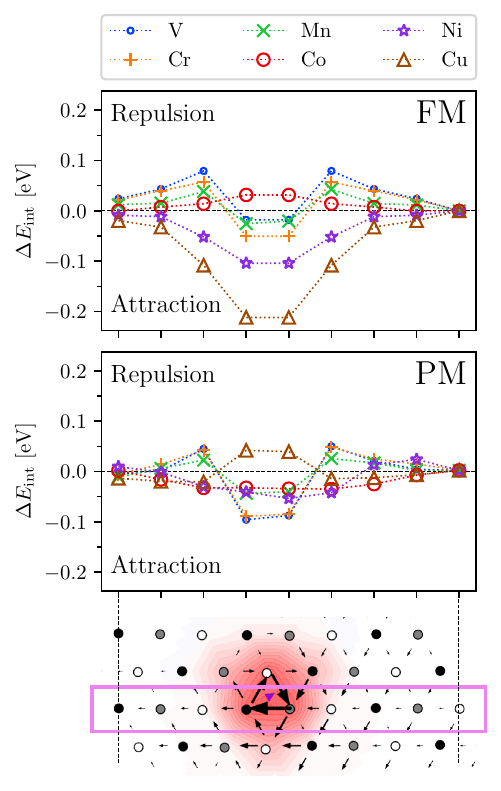}
   \caption{Energy profiles of the interaction energies across the dislocation core for \textit{3d}-elements (V, Cr, Mn, Co, Ni, Cu) in bcc Fe together with differential displacement maps and Nye tensor representation. The energies are calculated for both paramagnetic (PM) and ferromagnetic (FM) states using LSMS for a string of impurities in the $\Vec{b}$-direction.
   }
   \label{fig:fig_04}
\end{figure}

To analyze energy profiles around the core, we use GPU-accelerated 
LSMS for efficiency. As DLM is unavailable there, SWM is used. 
We examine a symmetric profile through the dislocation core, intersecting two atomic sites 
and extending toward the quadrupole center (see Fig.~1 in SM).
Results for magnetic states at $a_\mathrm{HT}$ 
(\cref{fig:fig_04}) include differential displacement maps and Nye tensor visualizations. 
$\Eint{}$ align with other methods except FM Mn, which exhibits magnetically frustrated sensitivity. 
In the PM state, SWM-LSMS and DLM-ELSGF yield qualitatively consistent outcomes.

Focusing on the shape of energy profiles, we can distinguish two main types: Monotonic (a single hill
or a valley at the core, e.g., Co, Ni, Cu in FM-Fe in \cref{fig:fig_04}) and
non-monotonic (peaks and valleys, e.g., V, Cr, Mn in FM-Fe \cref{fig:fig_04}).
The non-monotonic energy profiles for V, Cr, and Mn do not change when switching from the FM to the PM
state.
The monotonic profile for Co does not change its shape, but it flips from a repulsive (hill-like)
to attractive (valley-like) behavior.
The most interesting transformation happens to the profiles of Ni and Cu, whose shapes change
from the monotonic one in the FM state to the non-monotonic one in the PM state. 
Cu shows the strongest change in interaction tendency with 
temperature, linked to its solubility in $\alpha$-Fe: insoluble in the FM state at LT
but weakly soluble near the PM state at HT \cite{Gorbatov2013a}.

\begin{figure}
   \centering
   \includegraphics{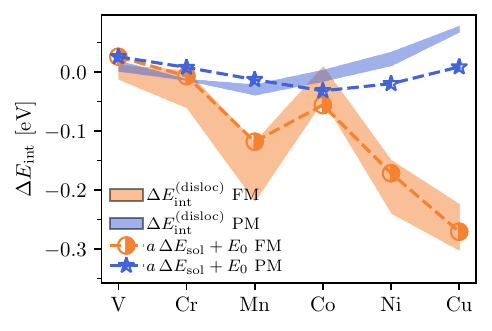}
   \caption{Lines: Interaction energy model from energy differences between fcc and bcc solution energies for \textit{3d}-elements (V, Cr, Mn, Co, Ni, Cu) in FM and PM states. Shaded areas: Range of interaction energies using different methods (see also \cref{fig:fig_01}b).
   }
   \label{fig:fig_05}
\end{figure}

Polyhedral template matching~\cite{Larsen2016} shows three core-adjacent 
sites are fcc-like for the compact core, while the rest are clearly bcc, as noted in Refs. \cite{Wang2024,aitken2024controlling}. 
We have examined the linear correlation between $\Eint$ 
and the difference in solution energies between fcc and bcc structures,
\begin{equation}
\Delta E_{\mathrm{sol}} = E_{\mathrm{sol}}^{(\mathrm{fcc})}(\mathrm{X} \to \mathrm{Fe}) -
    E_{\mathrm{sol}}^{(\mathrm{bcc})}(\mathrm{X} \to \mathrm{Fe}),
\end{equation}
where the solubility for a given structure is calculated using EMTO-CPA as 
\begin{equation}
    E_{\mathrm{sol}}(\mathrm{X} \to \mathrm{Fe}) =
        \left. \dfrac{\pd E(\mathrm{X}_c \mathrm{Fe}_{1-c})}{\pd c} \right|_{c \to 0}.
\end{equation}

Given above definitions, a model describing interaction trends can be written as
\begin{equation}
E_{\mathrm{seg}} = a \Delta E_{\mathrm{sol}} + E_0,
\end{equation}
where parameters $a$ and $E_0$ are determined by performing 
a least-squares fit on all data, for all elements, and for both FM and PM states.

The model in \cref{fig:fig_05} closely matches the true interaction energies for 
both magnetic states, suggesting that dislocation-impurity interactions are strongly 
influenced by nearest-neighbor coordination.

Apart from this, the fcc-like dislocation core contributes to 
the sensitivity of $\Eint{}$ 
among methods in the FM-state. In fcc Fe, impurities strongly perturb 
the magnetic state of surrounding atoms, leading to energy contributions from both 
impurities themselves and the magnetic moment \emph{reorganization}~\cite{ponomareva2014ab}. Qualitative trends remain, however, consistent.

To conclude, we have investigated the effect of the ferromagnetic-to-paramagnetic transition
in bulk $\alpha$-Fe on the interaction of \textit{3d}-metal impurities with a screw dislocation.
We have found that the magnetic state has a strong impact on the behavior of impurities, leading to
the inversion of the interaction for some of the elements (Ni, Cu) and strongly non-linear
behavior of the interaction for Co with temperature.
Especially large changes are observed for Cu, exhibiting strong attraction to the dislocation core
in the ferromagnetic state at low temperature, while becoming repulsive in the paramagnetic state
at high temperatures. Furthermore, this crossover is accompanied by significant changes in the shape of
the energy profile in the vicinity of the core. Similar, but weaker behavior is observed for Ni and Mn.
We have shown that the behavior of all impurities around a screw dislocation
significantly correlates with the differences between the solution energies of fcc and bcc structures of
respective elements.
The observed behavior of impurities can have implications 
on the plasticity of Fe-based alloys, especially
at temperatures around the Curie temperature, relevant for heat treatment. Future work 
could explore dislocations with DFT-accurate interatomic potentials~\cite{Hodapp2020}.

\section*{CRediT authorship contribution statement}

\textbf{F. M.}: 
    Conceptualization, 
    Methodology, 
    Software, 
    Formal analysis, 
    Data curation, 
    Writing – review \& editing,
    Writing - original draft
\textbf{L. R.}: 
    Funding acquisition, 
    Conceptualization, 
    Methodology, 
    Writing – review \& editing
\textbf{A. R.}: 
    Formal analysis,
    Software,
    Data curation
\textbf{S. G.}: 
    Software, 
    Methodology, 
    Writing – review \& editing
\textbf{M. E.}: 
    Software,
    Methodology,
    Writing – review \& editing
\textbf{O. P.}: 
    Software, 
    Supervision, 
    Funding acquisition,
    Writing – review \& editing,
    Writing - original draft

\section*{Declaration of competing interest}

The authors declare that they have no known competing financial
interests or personal relationships that could have appeared to influence
the work reported in this paper.

\section*{Acknowledgement}

F.M. would like to thank Y. Wang (PQI) and H. Xue (UTK) for discussions and hospitality during the stay at UTK and ORNL and T. Ruh for his technical support.
This work was supported by the Forschungsf\"orderungsgesellschaft (FFG)
project No. 878968 ``ADAMANT'', Austrian Science Fund (FWF) project No. P33491-N ``ReCALL'', COMET program IC-MPPE (project No 859480), and Austrian Marshall Plan Foundation
The COMET program is supported by the Austrian Federal Ministries for Climate Action, Environment, Energy, Mobility, Innovation and Technology (BMK) and for Digital and Economic Affairs (BMDW), represented by the Austrian research funding association (FFG), and the federal states of Styria, Upper Austria and Tyrol. This research used resources of the Oak Ridge Leadership Computing Facility, which is a DOE Office of Science User Facility supported under Contract DE-AC05-00OR22725, the Vienna Scientific Cluster (VSC-5) and the National Academic Infrastructure for Supercomputing in Sweden (NAISS), partially funded by the Swedish Research Council through grant agreement no. 2022-06725. The financial support by the Austrian Federal Ministry for Labour and Economy
and the National Foundation for Research, Technology and Development and the Christian Doppler
Research Association is gratefully acknowledged. We acknowledge AURELEO for awarding this project access to the LEONARDO supercomputer, owned by the EuroHPC Joint Undertaking, hosted by CINECA (Italy) and the LEONARDO consortium.

\bibliographystyle{custom}
\bibliography{references}

\clearpage
\newpage

\onecolumngrid
\vspace{2em}
\begin{center}
  {\large \bfseries
  Supplemental Materials: Inversion of Dislocation-Impurity Interactions in \texorpdfstring{$\alpha$}{α}-Fe under Magnetic State Changes}
\end{center}
\vspace{2em}
\twocolumngrid

\section{Interpolation of the interaction energy}

\newcommand{\LT}{\textsc{lt}}
\newcommand{\HT}{\textsc{ht}}

We would like to get the minimal consistent interpolation for the function $E(m, a)$,
given its values at four points $m = \{0, 1\}$, $a = \{ a_{\LT}, a_{\HT} \}$.
Assuming a linear dependence of the segregation energy on the lattice constant, we can write
\begin{align}
    E(m, a) = p(m) (a - a_{\LT}) + q(m),
\end{align}
with the boundary values for $p(m)$ and $q(m)$ defined as
\begin{align}
    p(1) = & \frac{ E(1, a_{\HT}) - E(1, a_{\LT}) }{ a_{\HT} - a_{\LT} }, &
            q(1) = & E(1, a_{\LT}), \\
    p(0) = & \frac{ E(0, a_{\HT}) - E(0, a_{\LT}) }{ a_{\HT} - a_{\LT} }, &
            q(0) = & E(0, a_{\LT}). \\
\end{align}

Assuming additionally that $p(m)$ and $q(m)$ are also linear functions of $m$, we get
\begin{align}
    p(m) = & m \left[ p(1) - p(0) \right] + p(0), \\
    q(m) = & m \left[ q(1) - q(0) \right] + q(0).
\end{align}

The obtained interpolation is bilinear in $a$ and $m$, with the coefficient in front
of $m \cdot a$ being
\begin{align}
   & \frac{ E(1, a_{\HT}) - E(1, a_{\LT}) - E(0, a_{\HT}) + E(0, a_{\LT}) }
        { a_{\HT} - a_{\LT} } \approx \\
   & \approx \frac{ E(1, a_{\HT}) - E(1, a_{\LT}) }
        { a_{\HT} - a_{\LT} },
\end{align}
where we make use of the observation that the segregation energy in the PM state is
only weakly dependent on $a$.

Given that in the FM state, $E(1, a_{\HT}) > E(1, a_{\LT})$, we conclude that
the coupling coefficient of $m \cdot a$ is always positive.

Another interesting aspect is the slope, $d E / d T$, just below the Curie point, $T_{C}$.
Since in this region, $\pd m / \pd T$, is very large in absolute value, we have for
$a \approx a_{\HT}$:
\begin{align}
    \frac{ d E }{ d T } = & \frac{ \pd E }{ \pd m } \frac{ \pd m }{ \pd T } +
        \frac{ \pd E }{ \pd a } \frac{ \pd a }{ \pd T }
        \approx \frac{ \pd E }{ \pd m } \frac{ \pd m }{ \pd T } \\
= &  \left[
        \Big( p(1) - p(0) \Big)( a - a_{\LT}) + q(1) - q(0)
    \right]
    \frac{ \pd m }{ \pd T } \\
\approx & \Big[ E(1, a_{\HT}) - E(0, a_{\HT}) \Big] \frac{ \pd m }{ \pd T }.
\end{align}

The temperature slope of the reduced magnetization is always negative and goes
to $-\infty$ as $T \to T_{C}$, therefore, the sign of $\pd E / \pd T$ is
determined by the difference of FM and PM energies at the HT lattice constant.
As one can see in Fig.~1 in the main text, this difference is positive for V, Cr,
and Co, resulting in a large negative slope for the energy at $T_{C}$ for these elements.
Since these elements have relatively weak interaction energies, the values at LT
and HT are similar, which forces the curve $E(T)$ to have a maximum at some temperature
below $T_{C}$. This is what we see in Fig.~3 of the main text.

\section{Dislocation cell setup}

The dislocations are modeled with the 
dipole approach with 135 atoms per 1 Burgers ($1b$) vector. The Burgers vector direction, 
$\Vec{b}$, is along the $\Vec{z}$. The cell vectors are given by $C = \{[1,1,\overline{2}], [7,\overline{2},\overline{5}], [\overline{1},\overline{1},\overline{1}] \}$. Plastic strain compensation $\varepsilon = - (\Vec{b} \otimes \Vec{A} + \Vec{A} \otimes \Vec{b}) / 2 \Omega$ was 
included. The vector $\Vec{A}$ is given by 
\begin{equation}
    \Vec{A} = \Vec{c} \times (\Vec{P}^{-} - \Vec{P}^{+}), 
\end{equation}
where $\Vec{c}$ is the lattice vector in Burgers vector direction and $\Vec{P}$ is the location 
vector of the dislocation core. The plastic strain compensation lead to a contribution 
in the $\Vec{z}$-component (component along $\Vec{b}$-direction) 
of the second lattice vector. 
The two location vectors, $\Vec{P}^{-}$ and $\Vec{P}^{+}$, are located at 
$\sfrac{1}{2} \, C_{11}, \sfrac{11}{27} \, C_{22}, 0$ and 
$a_{11}, \sfrac{10}{27} \, C_{22}, 0$, respectively, 
where $C_{ij}$ are the entries of the cell matrix $C$.

Canonical representation of the cell can be achieved by transforming $C \cdot U$, where $U = \{[1,1,\overline{2}], [1,\overline{1},\overline{0}], [\overline{1},\overline{1},\overline{1}] \}$. The cell vectors are fixed during ionic relaxation and only the ionic positions are moving.

\begin{figure}[!h]
\centering
\includegraphics[width=\columnwidth]{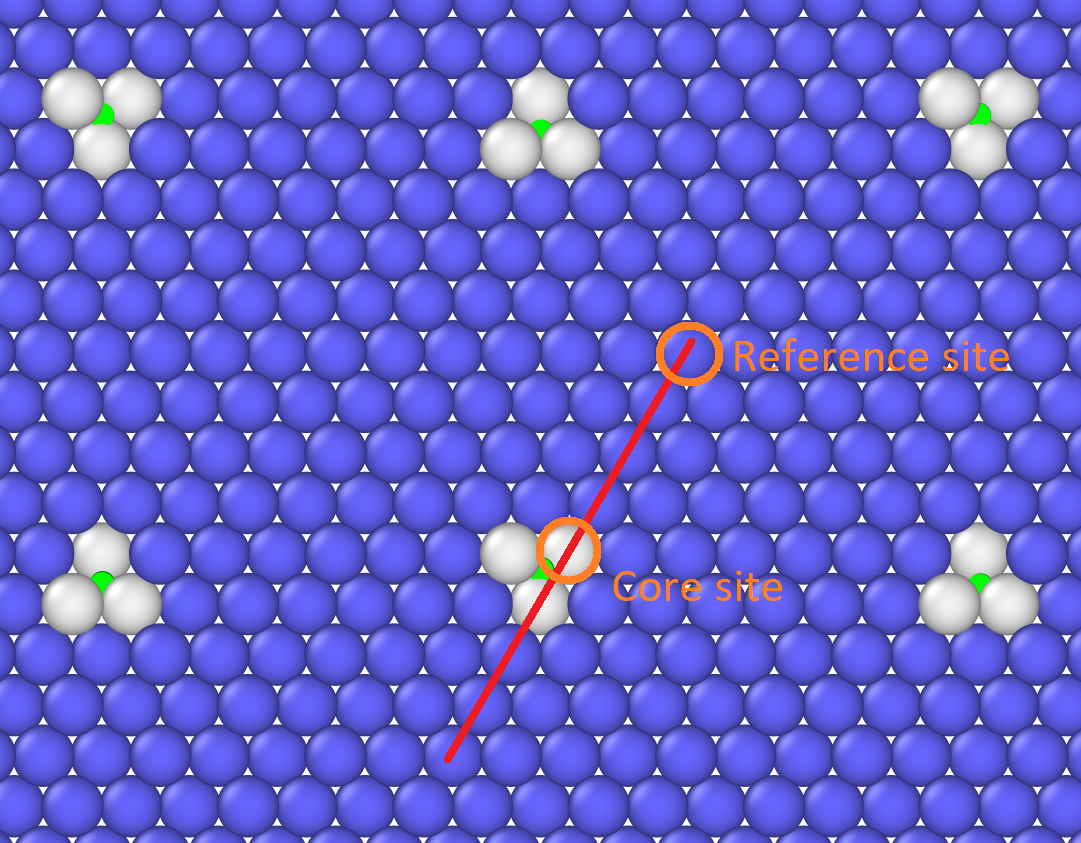}]
\caption{Dislocation dipole with highlighted sites where atoms were replaced. The core sites and the reference 
site, located six shells away from the core, are indicated. The red line marks row of sites that were 
replace for the dislocation-impurity interaction energy profile.}
\label{fig:fig_sm_02}
\end{figure}

\Cref{fig:fig_sm_02} shows the sites that were replaced for the energy profiles 
and the dislocation-impurity interactions calculations. The energy profile 
crosses the dislocation core and extends to the center region of the dislocation quadrupole. The 
center site and the reference site are marked by orange circles. The reference 
point is on the sixth shell away from the core following the definition of Rao 
(see Fig. 2 in Ref. \cite{Rao2021}). There, the first four shells surrounding 
the dislocation core are illustrated, along with the number of 
atomic sites belonging to each shell. The first two shells form a 
triangular pattern in the projection on the plane normal to the 
$\Vec{b}$-direction, while the subsequent shells have a hexagonal pattern.

In the main manuscript in Fig 4., the atomic sites in the profile that are directly adjacent to the three 
central core atoms belong to the third shell. From these sites outward, each adjacent shell 
is counted incrementally. Although the reference site appears five atomic shells away 
from the core, because it is the fifth site in the sequence counted from one of the 
dislocation core atoms, it actually belongs to the 
sixth shell when shells are defined based on geometric proximity.

As mentioned in the main text, we compare two setups for handling possible 
impurity-impurity interactions: (1) a chain (string) of impurity 
atoms along the z-axis (denoted as ``Str''), (2) a single impurity atom surrounded by Fe atoms along the z-axis 
(denoted as ``Imp''). 
In the CPA-based EMTO approach, the latter configuration is modeled as 
the dilute limit of the Fe-X alloy at a specific site. 

This means we define a CPA site with a 
low concentration ($1\,at.\,\%$) of the 
impurity element while maintaining a $1b$-height of the cell. In contrast, 
a $4b$-height is used in 
all other ``Imp'' computations. We performed VASP calculation to check 
the convergence of the cell height 
and found that $4b$ was sufficient for our quantities in question.

\section{Calculation setting}

\subsection{VASP}

For the VASP calculations, the \textbf{\_pv} pseudopotentials were 
used in all cases. A $\Gamma$-centered 
k-point grid of $1 \times 1 \times 16$ was used for the $1\Vec{b}$ cell, 
and $1 \times 1 \times 4$ for the $4\Vec{b}$ cell.

The following parameters were used in the \texttt{INCAR} file:

\begin{lstlisting}[basicstyle=\ttfamily\small]
ADDGRID     = TRUE
ALGO        = Fast
EDIFF       = 1e-7
EDIFFG      = -0.002
ENCUT       = 475
ISIF        = 2
ISMEAR      = 1
ISPIN       = 2
ISYM        = 2
LASPH       = False
LCHARG      = True
LMAXMIX     = 6
LREAL       = Auto
ROPT        = 1e-5 1e-5
LWAVE       = True
NELM        = 180
NELMDL      = -12
NELMIN      = 7
PREC        = Accurate
SIGMA       = 0.1
SMASS       = 0.5
SYMPREC     = 1.0e-5
AMIX        = 0.1
BMIX        = 0.0001
AMIX_MAG    = 0.4
BMIX_MAG    = 0.0001
LORBIT      = 11
EMAX        = 13.0
EMIN        = -10.0
NEDOS       = 11
\end{lstlisting}

\subsection{EMTO and ELSGF}

The EMTO and ELSGF calculations were performed using an \textit{spd}-basis 
set with a $1 \times 1 \times 19$ k-point grid. An elliptical contour with 14 energy points 
was employed for the integration. No Fermi contour was used for the temperature 
effects. To achieve faster convergence, the initial starting calculation 
was performed with fixed spin moments. The subsequently restarted calculations were performed with 
\emph{free} magnetic moments. The ELSGF calculations were performed with $\mathrm{LIZ} = 3$. Convergence 
was check for the LIZ and only for Mn significant difference was found between $\mathrm{LIZ} = 2$ and 
$\mathrm{LIZ} = 3$.

\subsection{LSMS}

The LSMS calculations were performed using a $\mathrm{LIZ} =16.0\,\mathrm{a.u.}$. The $l$-cutoff was chosen 
to be 3 and 31 grid points on the contour were used. For the spin-wave method, 8-point Monkhorst-Pack sampling was used to setup the spin waves.

\subsection{Models for Paramagnetic state}

We also use this section to compare various models for the high-temperature paramagnetic state.

One approach involves the single-site Disordered Local Moment (DLM) model within the Coherent Potential Approximation (CPA). Others are supercell methods that directly model disordered spins. 

Within the DLM picture, the magnetically disordered state can be described as a
pseudo-alloy of equal amounts of components on a CPA site with spin up and spin
down orientations of their local moments. 

In contrast, there are supercell method, which are centered on the 
idea that in the PM state, where local moments are completely 
disordered, the average spin–spin correlation functions approach zero. 

Given that interaction between spins parameters decay with distance in the case of a disordered state, 
it is assumed that interactions have a finite range. In an ideal Heisenberg model system, 
magnetic interactions are constant, making the satisfaction of these conditions independent 
of the specific magnetic structure. Thus, averaging can be performed over a proper set 
of magnetic systems. This set can include arbitrary magnetic structures.

Following this, the energy of the 
PM state in the Spin-Wave Method (SWM) is calculated by integrating the 
energies of planar spin spiral with wave vector $q$ over the Brillouin zone. This is done 
using the special point technique, which requires 
performing DFT calculations for a finite set of $q$-vectors, significantly 
reducing computational time.

\begin{figure}[!h]
   \centering
   \includegraphics{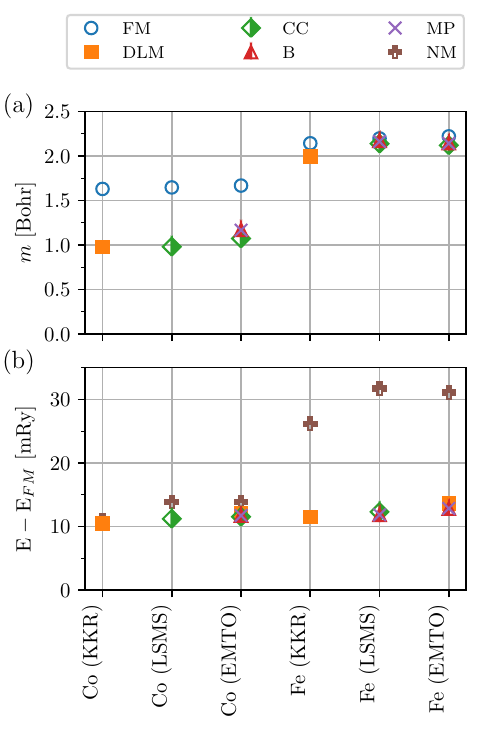}
   \caption{Comparison of magnetic moments and total energies from different DFT methods. FM refers to the ferromagnetic state, NM to the non-magnetic state, and DLM to the paramagnetic state with CPA. CC (Chadi-Cohen), B (Baldereschi), and MP (Monkhorst-Pack) refer to the paramagnetic state with the spin-wave method. (a) Average magnetic moments and (b) Total energy difference relative to the FM state of fcc Co and bcc Fe.
   }
   \label{fig:fig_sm_01}
\end{figure}

To assess the accuracy of the methodologies, we compare the energy 
difference between the PM/NM state and the FM state (see \cref{fig:fig_sm_01}). The PM state was calculated using 
spin-wave approach (SWM) and disordered local moment approach (DLM). The 
results show that the 8-point Monkhorst-Pack (MP) sampling, special point sampling 
with Chadi-Cohen (CC), and Balderschi (B) point sampling yield very similar values for 
SWM. Additionally, the overall energies from LSMS, KKR, and EMTO provide consistent results.

\end{document}